\documentclass[revtex4]{emulateapj}
\usepackage{epsf}
\usepackage{natbib}
\usepackage{graphicx}
\usepackage{float}
\usepackage{afterpage}
\usepackage{enumitem}
\usepackage{bm,bbm,amssymb,color, amsmath, color}
\usepackage{enumerate}
\usepackage{url}
\usepackage{enumitem}
\usepackage[section]{placeins}

\usepackage{mathrsfs}  
\DeclareMathOperator\erfc{erfc}

\usepackage{hyperref}

\begin{document}
\title{Diffusive Versus Free-Streaming Cosmic Ray Transport in Molecular Clouds}

\author{Kedron Silsbee$^1$, Alexei V. Ivlev$^1$}
\email[e-mail:~]{ksilsbee@mpe.mpg.de} \email[e-mail:~]{ivlev@mpe.mpg.de} \affiliation{$^1$Max-Planck-Institut f\"ur
Extraterrestrische Physik, 85748 Garching, Germany }

\begin{abstract}
Understanding the cosmic ray (CR) ionization rate is crucial in order to simulate the dynamics of, and interpret the chemical species observed in molecular clouds.  Calculating the CR ionization rate requires both accurate knowledge of the spectrum of MeV to GeV protons at the edge of the cloud as well as a model for the propagation of CRs into molecular clouds.  Some models for the propagation of CRs in molecular clouds assume the CRs to stream freely along magnetic field lines, while in others they propagate diffusively due to resonant scattering off of magnetic disturbances excited by MHD turbulence present in the medium.  We discuss the conditions under which CR diffusion can operate in a molecular cloud, calculate the local CR spectrum and ionization rate in both a free-streaming and diffusive propagation model, and highlight the different results from the two models.  We also apply these two models to the propagation through the ISM to obtain the spectrum seen by Voyager 1, and show that such a spectrum favors a diffusive propagation model.
\end{abstract}


\maketitle

\section{Introduction}
Cosmic rays (CRs) provide the dominant source of ionization in molecular clouds at visual extinctions greater than 1 to a few, depending on conditions, which corresponds to H$_2$ column depths greater than 1 to a few $\times 10^{21}$ cm$^{-2}$ \citep{McKee89, Keto08}.  They affect the gas-phase chemistry \citep{Dalgarno06}, chemistry that occurs in dust grains \citep{Shingledecker18}, and contribute heating to cold cores of molecular clouds \citep{Glassgold12, Galli15}.
\par
Many widely adopted models for the propagation of CRs in molecular clouds \citep{Padovani09, Padovani18} assume that CRs stream freely along magnetic field lines.  Under this assumption, \citet{Padovani18} showed that the CR ionization rate at a point in the cloud is a function of the {\it effective} column density $N$ to that point, integrated along the magnetic field lines.  

\par
There is also discussion in the literature of diffusive CR propagation.  Turbulence can excite MHD waves that scatter the CRs' pitch angles \citep{Kulsrud69}, leading to spatial diffusion.  This turbulence can arise from the anisotropy in the CR distribution function which arises near the cloud in response to CR absorption in the cloud center.  The role of such turbulence has been discussed in several works \citep[e.g.,][]{Skilling76, Morlino15, Ivlev18}.  However, it was found that for clouds with $N \lesssim 10^{23}$ cm$^{-2}$, the effect of such self-generated turbulence on CR penetration is only marginally significant \citep{Dogiel18}. 
\par
Diffusive CR transport can also occur due to pre-existing turbulence.  \citet{Schlickeiser16} calculated the CR spectrum and resulting ionization rate, assuming an energy-dependent diffusion coefficient derived in \citet{Schlickeiser10} and using the CR spectrum derived from the Voyager 1 results.  
\par
There is increasing evidence for both substantial variability from object to object, as well as a steep dependence of the ionization rate on the column density of the cloud.  Estimates of the primary CR ionization rate per hydrogen atom $\zeta_p$ from observations of OH$^+$ in low-density clouds give values ranging from $3.9 \times 10^{16}$ s$^{-1}$ up to $1.6 \times 10^{-15}$ s$^{-1}$ \citep{Bacalla18}.  A paper by \citet{Neufeld17} uses H$_3^+$ observations in different clouds with known column densities to determine $\zeta_p$ as a function of column density $N$.  While there is significant uncertainty in the results, they suggest a quite steep dependence of $\zeta_p$ on $N$.  Finally, \citet{Galli15} argue that the temperature and molecular abundance profile in the center of the starless core L1544 is best fit by $\zeta_p \sim 10^{-17}$ s$^{-1}$ or even lower.
\par

In this paper we discuss the possible role of pre-existing MHD turbulence.  Envelopes of molecular clouds are thought to be turbulent environments.  There is some uncertainty,  however,  as to whether the MHD waves associated with the turbulence would have sufficient energy at small enough scales to be resonant with the CRs responsible for the majority of the ionization.  Radio scintillation observations \citep[e.g.,][]{Armstrong95} suggest that turbulence in the ISM extends to scales at least as small as $10^{10}$ cm, comparable to the gyroradius of a sub-relativistic proton, but it is not clear that this result is relevant to molecular clouds.  As we show in Section \ref{molCloudIon}, assuming diffusive propagation of CRs into molecular clouds to take place, it would create a steep dependence of the ionization rate on column density although this is only likely up to column densities of $\sim 10^{21}$ cm$^{-2}$ under conditions appropriate for local molecular clouds.
\par 
The CR ionization rate depends on both the propagation model for CRs, and on their spectrum at the edge of the cloud.  Ionization at column densities in the range of $10^{20} - 10^{23}$ cm$^{-2}$ is dominated by CR protons with energies from 1 MeV to 1 GeV.  Unfortunately, the spectrum of such particles cannot be measured accurately from near Earth because they are largely excluded by the solar wind \citep{Potgieter13}.  The Voyager 1 probe has measured the spectrum of Galactic CRs down to 3 MeV \citep{Cummings16}.  However, the magnetic field direction measured by the probe has not changed, as it would be expected to if Voyager were really in a region of space beyond the influence of solar modulation \citep{Gloeckler15}.  Furthermore, \citet{Padovani18} and \citet{Phan18} noted that the proton and electron spectra from Voyager were too low by about a factor of 10 to explain the values of $\zeta_p$ observed in nearby molecular clouds.  For these reasons, there is still considerable uncertainty about the density of low-energy CRs impinging on molecular clouds. 
\par
Models for the acceleration of CRs in shocks suggest that they should act as power-law source functions for CRs [see e.g. \citet{Drury83}].  This is very different from the Voyager spectrum, which shows a broad turnover around 30 MeV.  Recently, there has been work to reproduce the spectrum seen by Voyager with the code GALPROP, using a complicated model including diffusion, advection, reacceleration, adiabatic momentum gain and loss and several energy loss processes \citep{Bischoff19}.  They are able to well reproduce the spectrum seen by Voyager.  In this paper we consider a simpler model, which includes the effect of a shell of high-density material surrounding the local bubble with magnetic field nearly parallel to the shell \citep{Alves18}.  In section \ref{voyager}, we look at the spectrum of CRs that would be seen by Voyager after propagation through this shell.  We find that diffusive propagation within this thin dense region could attenuate the power-law source spectrum of low-energy CRs, to produce something qualitatively resembling the Voyager spectrum.  The value of the column density required for such attenuation is much more reasonable than that predicted by a model of free-streaming propagation.

\section{Diffusive propagation Model}
In this section we calculate the CR spectrum as a function of $N$, assuming CRs propagate diffusively through an attenuating column.  The propagation is modelled as occurring due to a combination of diffusion along the magnetic field and energy losses due to ionization. 
\par
 Following \citet{Skilling75}, we use a simplified expression for the CR diffusion coefficient $D$ due to the presence of weak MHD turbulence:
\begin{equation}
D(E) = \frac{vB^2}{6\pi^2\mu_*k^2W},
\label{Dexp}
\end{equation}
where $v$ is the speed of the CR particle, $B$ the magnetic field strength, and $\mu_*$ is the  ``effective" cosine of the resonant pitch angle.  Furthermore, $k$ is the wavenumber of the resonant MHD wave, and $W(k)$ is the spectral energy density of MHD waves.  $k$ in Equation \eqref{Dexp} is related to $E$ via the cyclotron resonance condition, which is expressed in terms of $\mu_*$ as
\begin{equation}
k_{\rm res}(E) = \frac{m \Omega}{\mu_*p(E)}.
\label{kres}
\end{equation}
Here, $m$ is the mass of the CR particle, $\Omega$ is the CR gyrofrequency given by 
\begin{equation}
\Omega = \frac{eB}{mc},
\end{equation}
$c$ is the speed of light and $e$ the electron charge.  We assume that $W$ is equal to the ion kinetic energy density of the weak turbulence:
\begin{equation}
W(k) = \frac{\rho_{\rm ion} v^2_{\rm turb}(k)}{2k},
\end{equation}
where $\rho_{\rm ion}$ is the mass density of ions in the medium.  We assume a power law spectrum of the turbulence, such that
\begin{equation}
v_{\rm turb} = v_* \left(\frac{k_*}{k}\right)^\lambda,
\label{vturb}
\end{equation}
where $\lambda$ is 1/3 for a Kolmogorov turbulent spectrum, and 1/4 for a Kraichnan spectrum.  Putting all of this together, we find $D \propto E^{1-\lambda} \rho_{\rm ion}^{-1}$.
\par
Let us consider the flux of particles entering the column from the source along a field line.  Let $s$ be the coordinate of distance along this line.  The source is located at $s = 0$, and $s$ increases with increasing column away from the source.  The transport equation for $n(E, s)$, the differential density of particles with energy $E$ at position $s$, considering spatial diffusion and energy losses is given by
\begin{equation}
\frac{\partial n}{\partial t} = \frac{\partial}{\partial s}\left(D \frac{dn}{ds}\right) -  \frac{\partial}{\partial E}\left(\frac{dE}{dt}n\right),
\label{transport}
\end{equation}
where $dE/dt$ is given in terms of the loss function $L$ as 
\begin{equation}
\frac{dE}{dt}  = -L(E)n_g v(E).
\label{dedt}
\end{equation}
Here $n_g$ is the density of hydrogen atoms $n_g = n(H) + 2n(H_2)$.  We are searching for a steady-state distribution of $n$, and set $\partial n/ \partial t = 0$.  Noting that $dN = n_g ds$, and using Equation \eqref{dedt}, Equation \eqref{transport} can be written as 
\begin{equation}
X(E) \frac{\partial F}{\partial E} + \frac{\partial ^2F}{\partial N^2}= 0,
\label{middleStep}
\end{equation}
where 
\begin{equation}
F(E, N) = n(E, N)v(E)L(E),
\end{equation}
and $X(E) =L(E)v(E)/\Delta(E)$, with the rescaled diffusion coefficient $\Delta$ given by 
\begin{equation}
\Delta(E) = n_g(N) D(E, N).
\end{equation}
  Note that we are able to write Equation \eqref{transport} in this form because we have assumed that both the turbulent velocity spectrum and the magnetic field strength are independent of $N$, and that the ionization fraction is constant, so $\rho_{\rm ion} \propto n_g$.  Under these assumptions, $\Delta$ is a function just of $E$ (see discussion in Section \ref{rangeOfApp}).
 \par
We then let 
\begin{equation}
 T= -\int_0^E\frac{dE'}{X(E')},
 \label{Meq}
\end{equation}
where $-T$ corresponds to the amount of diffusion undergone by a particle of energy $E$ during the time it loses all of its energy.  Then, Equation \eqref{middleStep} becomes a linear diffusion equation 
\begin{equation}
 \frac{\partial F}{\partial T} = \frac{\partial ^2F}{\partial N^2},
\label{diffEq}
\end{equation}
with $T(E)$ being the pseudo-time.  This allows the solution in a general form of the problem where $F(T, N)$ at the boundary $N = 0$ is a given function of pseudo-time \citep{Landau59}:
\begin{equation}
F(T, N) = \int_{-\infty}^T \frac{N\exp{(-\frac{N^2}{4(T-T_i)})}}{\sqrt{4\pi (T-T_i)^3}} F_i(T_i)dT_i,
\label{Fsol}
\end{equation}
where $F_i(T_i)$ is determined by the spectrum $j_i$ of CRs on the outside of the cloud as a function of their initial energy $E_i$, with $T_i = T(E_i)$.
\section{ionization rate in the envelopes of molecular clouds}
The primary CR ionization rate of H$_2$, $\zeta_{\rm H_2}$, can be calculated using the relation 
\begin{equation}
\zeta_{\rm H_2}(N) = \int_0^\infty j(E, N) \sigma_{\rm H_2}(E) dE,
\label{genericionization}
\end{equation}
where $\sigma_{\rm H_2}$ is the ionization cross section for molecular hydrogen and $j(E, N) = n(E, N)v(E)$.  
\par
The loss function for protons is well approximated by a power law over the range of energies from $10^5$ to $10^9$ eV (relevant for ionization in molecular clouds) as:
\begin{equation}
L(E) = L_0\left(\frac{E}{E_0}\right)^{-d},
\label{Leq}
\end{equation}
with $L_0 = 1.27\times 10^{-15}$ eV cm$^2$, $E_0 = 1$ MeV, and $d = 0.82$ \citep{Padovani18}.  The use of approximation \eqref{Leq} facilitates the analysis below by simplifying the calculations significantly.  In \citet{Padovani18}, they assumed all the hydrogen to be in molecular form and used a column density which was the number of {\it particles} per unit area.  In this paper, since we are dealing with lower column densities where not all the hydrogen need be molecular, we define the column density $N$ as the number of {\it hydrogen atoms} per unit area.  This means that at a given mass surface density, our column density is higher than that in \citet{Padovani18} by a factor of 1.67, which we have taken into account in the value of $L_0$ in Equation \eqref{Leq}.  Using Equations \eqref{Dexp} through \eqref{vturb}, we obtain
\begin{equation}
\Delta(E) = \Delta_0 \left(\frac{E}{E_0}\right)^{1-\lambda},
\label{simpleD}
\end{equation}
where the value of $\Delta_0$ is discussed in Section \ref{rangeOfApp}.  Using Equations \eqref{Meq} and \eqref{Leq}, we can write $T$ as 
\begin{equation}
T = -\frac{1}{4}N_{0d}^2\left(\frac{E}{E_0}\right)^\alpha
\label{analT}
\end{equation}
where 
\begin{equation}
N_{0d} = \sqrt{\frac{4\Delta_0E_0}{\alpha v_0L_0}}
\end{equation}
is the characteristic column density necessary to attenuate a particle with energy $E_0$ for diffusive transport, $\alpha = 3/2 + d - \lambda$, and $v_0 = \sqrt{2E_0/m}$.  For protons in the range from $10^5$ eV to $5 \cdot 10^8$ eV, the ratio between the loss function and the H$_2$ ionization cross section is nearly constant \citep{Padovani18}.  Using the expression for the ionization cross section given in \citet{Rudd85}, and the loss function in Equation \eqref{Leq}, we determine this ratio to be approximately $\epsilon = 37$ eV.  Note that this corresponds to an energy lost per H$_2$ ionization event of approximately 62 eV, which is reduced by the ratio of the hydrogen number density to the particle number density.  Thus, we can write $\sigma_{\rm H_2} = L/\epsilon$, and Equation \eqref{genericionization} becomes
\begin{equation}
\zeta_{\rm H_2}(N) =  \int_0^\infty \frac{F(E, N)}{\epsilon} dE.
\label{zetaR}
\end{equation}
If we assume the following initial CR spectrum, on the outside of the cloud:
\begin{equation}
j_i(E) = j_0\left(\frac{E}{E_0}\right)^{-a},
\label{initJ}
\end{equation}
then, using Equation \eqref{analT}, one can express Equation \eqref{Fsol} in terms of $E$ as 
\begin{equation}
 j(E, N) = j_i(E) \int_0^1 \erfc{\left(\frac{N/N_{0d}}{\sqrt{\left(E/E_0\right)^\alpha (x^{-\frac{\alpha}{a+d}}-1)}}\right)} dx.
 \label{Jeq}
\end{equation}
For the local spectrum \eqref{Jeq}, we can, after some manipulation, write the ionization rate as 
\begin{equation}
\zeta_{\rm H_2}(N) = \frac{j_0L_0E_0 I_d}{\sqrt{\pi} \epsilon } \left(\frac{N}{N_{0d}}\right)^{-\gamma_d},
\label{analDiffusiveZeta}
\end{equation}
where
\begin{equation}
I_d = \int_0^\infty x^{\frac{\gamma_d-1}{2}} dx  \int_0^1 e^{-\frac{x}{1-y^\alpha}} \left(1-y^\alpha \right)^{-3/2} dy,
\label{Ieq}
\end{equation}
and
\begin{equation}
\gamma_d = \frac{4(a+d-1)}{3+2d-2\lambda}.
\end{equation}
For $\lambda = 1/3$ or $1/4$, and $a$ in the range [0.5, 2], the formula $I_d = (1.73-\lambda/3)/(a+d-1)$ is accurate to within 8\%; $I_d$ is convergent as long as $\gamma_d > 0$.

 \subsection{ionization Rate for Free-Streaming CRs}
 In the free-streaming approximation, we replace the diffusive flux $-D dn/ds$ in Equation \eqref{transport} with the free-streaming flux $\mu j$, where $\mu$ is the cosine of the pitch angle.  In this case, we can directly relate the initial energy $E_i$ to $E$, via $N$ and $\mu$, using the loss function:
 \begin{equation}
 N = \mu \int_E^{E_i} \frac{dE}{L(E)}.
 \label{genericN}
 \end{equation}
 Then, in the continuously slowing-down approximation \citep{Padovani09}, we find that 
 \begin{equation}
 F(E, N, \mu) = F_i(E_i, \mu),
 \label{CDSAj}
 \end{equation}
where $F_i$ is determined by the initial spectrum $j_i$.  Note that at the low densities relevant for our problem  the magnetic field strength can be assumed to be constant \citep{Crutcher12}.  In our analytic model, using Equation \eqref{Leq}, and \eqref{genericN} we can write 
\begin{equation}
E_i = \left(E^{1+d} + \frac{N}{\mu N_{0f}}E_0^{1+d}\right)^{1/(1+d)},
\label{Ei}
\end{equation}
where 
\begin{equation}
N_{0f} = \frac{E_0}{(1+d)L_0}
\end{equation}
is the characteristic column density necessary to attenuate a particle of energy $E_0$ for free-streaming transport.  Then the ionization rate is 
\begin{equation}
\zeta_{\rm H_2}(N) = \int_0^1 d\mu \int_0^\infty \frac{F(E, N, \mu)}{\epsilon} dE.
\label{genericStreamingZeta}
\end{equation}
Assuming an initial spectrum $j_i(E)$ given by Equation \eqref{initJ}, the local spectrum is
\begin{equation}
j(E, N, \mu) = j_i(E) \left[1 + \frac{N}{\mu N_{0f}} \left(\frac{E_0}{E}\right)^{1+d}\right]^{-\frac{a+d}{1+d}}.
\label{CDSAj}
\end{equation}
Using Equations \eqref{genericStreamingZeta} and \eqref{CDSAj} we can then write 
\begin{equation}
\zeta_{\rm H_2}(N) = \frac{(1+d)}{(a+2d)} \frac{j_0L_0E_0 I_f}{\epsilon} \left(\frac{N}{N_{0f}}\right)^{-\gamma_f},
\label{analStreamingZeta}
\end{equation}
where
\begin{equation}
I_f = \int_0^\infty  \left(x^{1+d} + 1\right)^{-\frac{a+d}{1+d}}dx
\end{equation}
and
\begin{equation}
\gamma_f = \frac{a+d-1}{1+d} \equiv \frac{3 + 2d - 2\lambda}{4(1+d)} \gamma_d.
\end{equation}
In the range from $a = [0.5, 2]$, the approximation $I_f = 1.07/(a+d-1) + 0.42$ is accurate to within 2\%.
\subsection{Diffusion Constant and Range of Applicability}
\label{rangeOfApp}
The diffusion approximation is only appropriate at column densities such that the particle has lost the memory of the pitch angle with which it entered the cloud.  Diffusion is approximately equivalent to a random walk at velocity $v$ with step length (in column density) $\delta N(E) = 3\Delta(E)/v$.  Thus, for a particle with energy $E$, the transition from free-streaming to diffusive propagation should occur roughly at the column $N$ such that $N \sim \delta N(E)$.  Using Equation \eqref{simpleD}, and keeping in mind that the particles responsible for the bulk of the ionization are sub-relativistic, we find 
\begin{equation}
\delta N(E) = \frac{3\Delta_0}{v_0} \left(\frac{E}{E_0}\right)^{1/2 - \lambda}.
\label{stepLength}
\end{equation}
The particles dominating the ionization at column $N$ are those whose stopping range $N_{\rm st}(E)$ is comparable to $N$.  The stopping range is calculated using Equation \eqref{genericN} as
\begin{equation}
N_{\rm st}(E) = N_{0f} \left(\frac{E}{E_0}\right)^{1+d}.
\label{range}
\end{equation}
Solving Equation \eqref{range} for $E$, and plugging the result into Equation \eqref{stepLength}, we find that the condition $N \gtrsim \delta N$ is appropriate for column densities
\begin{equation}
N \gtrsim \frac{3\Delta_0}{v_0} \left(\frac{3\Delta_0}{v_0N_{0f}}\right)^\frac{1 - 2\lambda}{1 + 2d + 2\lambda}.
\label{appropriateColumn}
\end{equation}
We estimate the diffusion constant $\Delta_0$, entering Equation \eqref{simpleD}, based on an assumed slope of the turbulent power spectrum.  We normalise the power spectrum based on observations of turbulent velocities at large scales.  To estimate $v_*$ and $k_*$ in Equation \eqref{vturb}, we assume a turbulent velocity of 1 km s$^{-1}$ at a scale of 1 parsec.  This requires a major extrapolation, and it is possible that the turbulence is damped by ion neutral friction at intermediate scales (see \citet{Soler13} for a thorough discussion of which MHD modes propagate at the scales of ion-neutral decoupling).  Despite these uncertainties, we point to \citet{Armstrong95} as evidence that a Kolmogorov power spectrum over a very wide range of $k$ is possible in the ISM.  We further assume that the ionization is dominated by singly-ionised carbon, with an abundance relative to hydrogen of $1.5 \times 10^{-4}$ \citep{Gerin15}.  We take $B = 3\, \mu G$, consistent with the results from \citet{Crutcher12}, and set $\mu_* = 2/3$.  
\par
Assuming a Kolmogorov turbulent spectrum between 1 parsec and the range of interest ($\lambda = 1/3$), Equations \eqref{Dexp} through \eqref{vturb} give $\Delta_0 = 4.2 \times 10^{28}$ cm$^{-1}$s$^{-1}$; for a Kraichnan spectrum ($\lambda = 1/4$), we find $\Delta_0 = 2.6 \times 10^{27}$ cm$^{-1}$ s$^{-1}$.  Plugging these values of $\Delta_0$ into Equation \eqref{appropriateColumn}, we find that the diffusion approximation is appropriate for $N \gtrsim 8 \times 10^{19}$ cm$^{-2}$ for the Kolmogorov spectrum, and $\gtrsim 3 \times 10^{18}$ cm$^{-2}$ for the Kraichnan spectrum.  We note that there is some observational evidence \citep{Heyer15} for a steeper turbulent spectrum of $\lambda = 0.5$ at spatial scales much larger than those resonant with sub-relativistic CRs.  However at these scales, the turbulence is supersonic, so the spectrum is not governed by the same physics. 
\par
At a certain higher column density, the ion density is expected to drop dramatically when there are no longer sufficient UV photons to keep carbon ionised.  This depends on the strength of the UV field near the cloud, as well as on the assumed properties of the medium \citep{Hollenbach99}.  Based on the work of \citet{Keto08}, we assume the transition to take place at $N_{\rm tran} \approx 2 \times 10^{21}$ cm$^{-2}$, though we note that \citet{Neufeld17} find a very sharp drop in C$^{+}$ abundance near a column density of $6 \times 10^{20}$ cm$^{-2}$.  For column densities greater than $N_{\rm tran}$, the ion density is expected to drop by a factor of $\sim$ 100, depending on $\zeta_{\rm H_2}$ \citep{Neufeld17}, leading to the proportional increase in $\Delta_0$.  Then it appears unlikely that the turbulence would be strong enough to greatly influence the CR propagation.  In environments with higher CR fluxes or more incident UV radiation than assumed by \citet{Neufeld17}, this boundary may be moved to higher column density.  Specifically, \citet{Neufeld17} find that if $\zeta_{\rm H_2}/n_H > 1.2 \times 10^{-17} {\rm cm}^{3} {\rm s}^{-1}$, then the ionization fraction remains greater than $10^{-4}$ to a column density of $\sim 10^{22}$ cm$^{-2}$.  Such conditions may be found near the Galactic center \citep{Petit16}.

\section{results for a model interstellar spectrum}
\citet{Padovani18} propose an interstellar CR spectrum of the following model form 
\begin{equation}
j(E) = C \frac{E^\delta}{(E + E_t)^\beta}\, {\rm eV}^{-1} {\rm cm}^{-2} \rm{s}^{-1}. 
\label{genericSpectrum}
\end{equation} 
The high-energy slope of this function, $\delta - \beta$, is well determined \citep[e.g.,][]{Aguilar14, Aguilar15}, while the low-end slope $\delta$ is uncertain.  \citet{Ivlev15} argue that the spectrum determined by Voyager \citep{Cummings16}, represents a lower bound on the interstellar proton spectrum, and they estimate an upper bound based on observed ionization rates in nearby clouds, in which $C = 3.0 \times 10^{16}$, $E_t = 650$ MeV, $\delta = -0.8$, and $\beta = 1.9$. 
\subsection{Ionization in molecular cloud envelopes}
\label{molCloudIon}
\begin{figure}[htp]
\centering
\includegraphics[width=1.03\columnwidth]{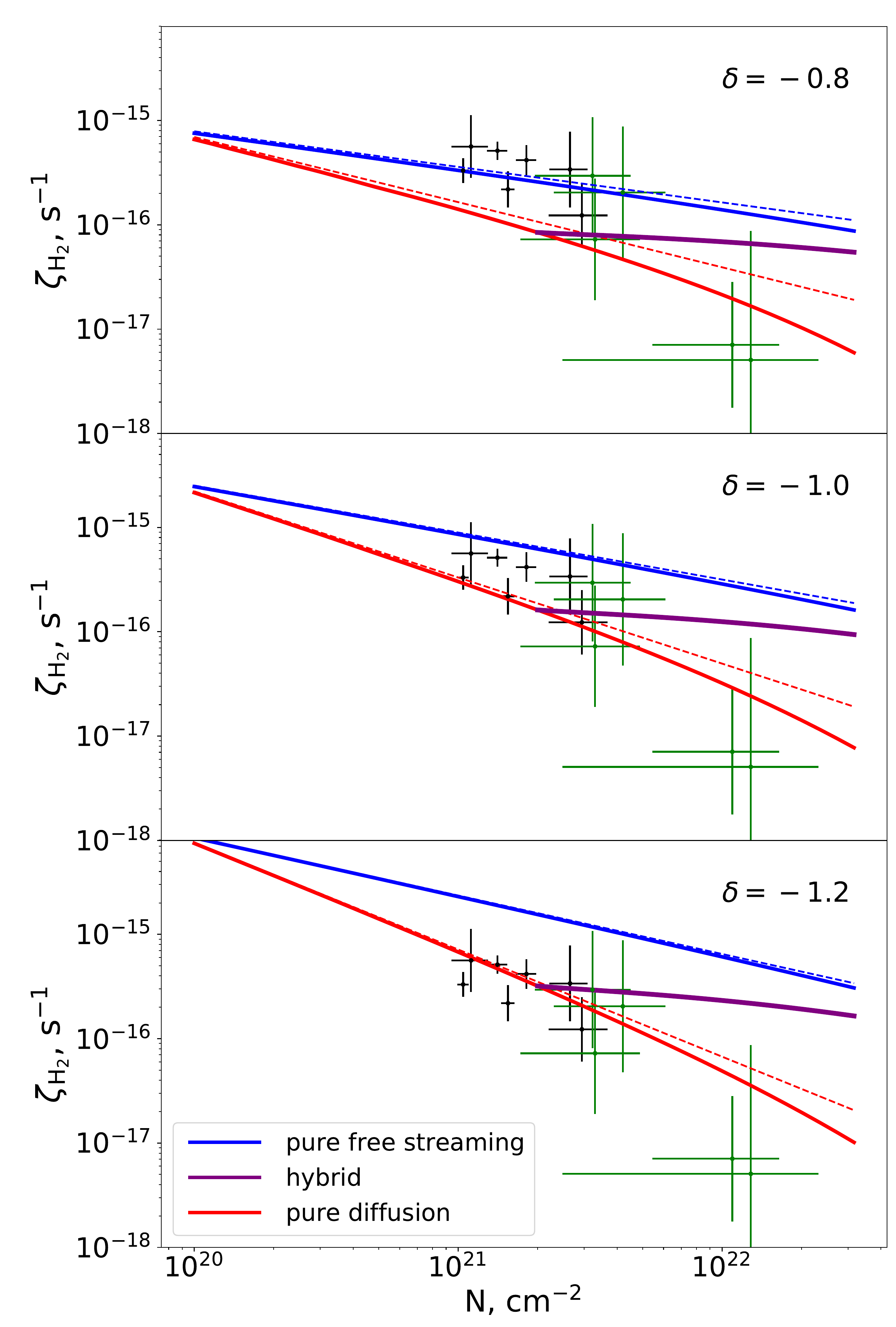}
\caption{Primary ionization rate of H$_2$, $\zeta_{\rm H_2}$, as a function of column density of hydrogen atoms $N$ for our three different propagation models, as indicated in the legend.  Different panels correspond to different values of $\delta$ in the assumed initial spectrum [see Equation \eqref{genericSpectrum}].  The points and error bars are estimated from Figure 6 of  \citet{Neufeld17}.  Black points are those for which the the H$_2$ column density has been measured directly \citep{Neufeld17}.  The solid lines represent the results for CR spectrum \eqref{genericSpectrum}, whereas the dashed lines are for the power-law spectrum in Equation \eqref{initJ}.  Details of the different propagation models are discussed in Section \ref{molCloudIon}.  Note that the column density displayed here is a factor 1.67 larger than that in \citet{Padovani18}}
\label{zetaFig}
\end{figure}

We use the spectrum described by Equation \eqref{genericSpectrum}, truncated at 3 GeV, to calculate $\zeta_{\rm H_2}(N)$ for three different propagation models described below.  The results are plotted in Figure \ref{zetaFig}.  In all cases, when calculating the ionization rate, we integrated Equation \eqref{zetaR} or \eqref{genericStreamingZeta} as appropriate from 10 KeV to 1 GeV.  We vary $\delta$ as labelled in the panels, using $C = 3.0 \times 10^{16}$, $E_t = 650$ MeV, and $\delta - \beta = -2.7$.  
\par
The blue curve assumes pure free-streaming propagation, in which $\zeta_{\rm H_2}(N)$ was determined using Equation \eqref{genericStreamingZeta}.  
\par
The purple curve represents the hybrid model, which assumes that CRs propagate diffusively until the column depth $N_{\rm tran} = 2 \times 10^{21}$ cm$^{-2}$ (such that carbon is no longer ionised) and then stream freely.  The left-hand part of the curve is described by Equation \eqref{zetaR} where $F(E, N)$ is given by Equation \eqref{Fsol}, with $\Delta_0$ evaluated in Section \ref{rangeOfApp} for Kolmogorov turbulence.  The hybrid model results in a region of nearly flat $\zeta_{\rm H_2}(N)$ at $N \gtrsim N_{\rm tran}$, where the spectrum is dominated by particles with $N_{\rm st} \gg N_{\rm tran}$.  Therefore, further attenuation has little effect until the column penetrated in the free-streaming region is comparable to the actual column passed through by the particles as they propagated diffusively.   
\par
The red curve assumes pure diffusive propagation for the entire column, ignoring the expected sharp decrease in $\rho_{\rm ion}$ that occurs around $N_{\rm tran}$.  This represents a lower bound on $\zeta_{\rm H_2}$, but such a curve is probably unrealistic unless there is some process (anomalously high UV field, or anomalously high $\zeta_{\rm H_2}$) that keeps a higher ionization fraction deeper within the cloud. 
\par
 Finally, the dashed red and dashed blue lines are the corresponding analytic approximations (given by Equations \eqref{analDiffusiveZeta} and \eqref{analStreamingZeta} respectively), assuming a spectrum given by Equation \eqref{initJ}, with $a = -\delta$ and $j_0 = C E_0^\delta/E_t^\beta$.  This spectrum coincides with that in Equation \eqref{genericSpectrum} at lower energies.  
 \par
The data points and error bars in Figure \ref{zetaFig} are taken from Figure 6 of \citet{Neufeld17}\footnote{\citet{Neufeld17} plot $\zeta_p$, the primary ionization rate per hydrogen, which they assume to be 1/2.3 times the total ionization rate $\zeta_t$ (including secondary ionizations) per H$_2$.  Taking a ratio $\zeta_t/\zeta_{\rm H_2} = 1.7$ \citep{Glassgold12}, we find that we must shift the points from \citet{Neufeld17} upwards by a factor of 1.4.}, assuming one magnitude of visual extinction to be equivalent to a column density of $1.9 \times 10^{21}$ cm$^{-3}$.  The H$_2$ column density for the black points was measured directly, whereas for the green points it was inferred from meausrements of CH or the reddening.  The spectrum in the top panel, corresponding to $\delta = -0.8$ in Equation \eqref{genericSpectrum}, was constructed by \citet{Padovani18} so that the free-streaming model passes through the points.  For the other models, this spectrum yields curves which are too low.  In the middle panel we plot the resulting ionization rate if $\delta$ is changed to 1.0.  The low-energy slope of the resulting spectrum corresponds to the spectrum of particles produced in strong shocks \citep{Drury83}.  Finally, in the bottom panel, we consider a steeper low-energy slope of $\delta = 1.2$.  In this case, the diffusive model provides the best fit.  We also point out that the slope for $\zeta_{\rm H_2}(N)$ obtained from Neufeld ($1.05 \pm 0.36$) is fit better by the diffusive model with $\delta = 1.2$, (which has a slope of 1.1 at $N = 10^{21}$ cm$^{-2}$), compared with the free-streaming model with $\delta = 0.8$, (which has a slope of $-0.4$ at at $N = 10^{21}$ cm$^{-2}$).  This argument would seem to favor the diffusive model.  However, as is clear from the magnitude of the error bars, the slope suggested by \citet{Neufeld17} is rather uncertain.  
\par
\subsection{Voyager Spectrum}
\label{voyager}
\begin{figure}[htp]
\centering
\includegraphics[width=1.03\columnwidth]{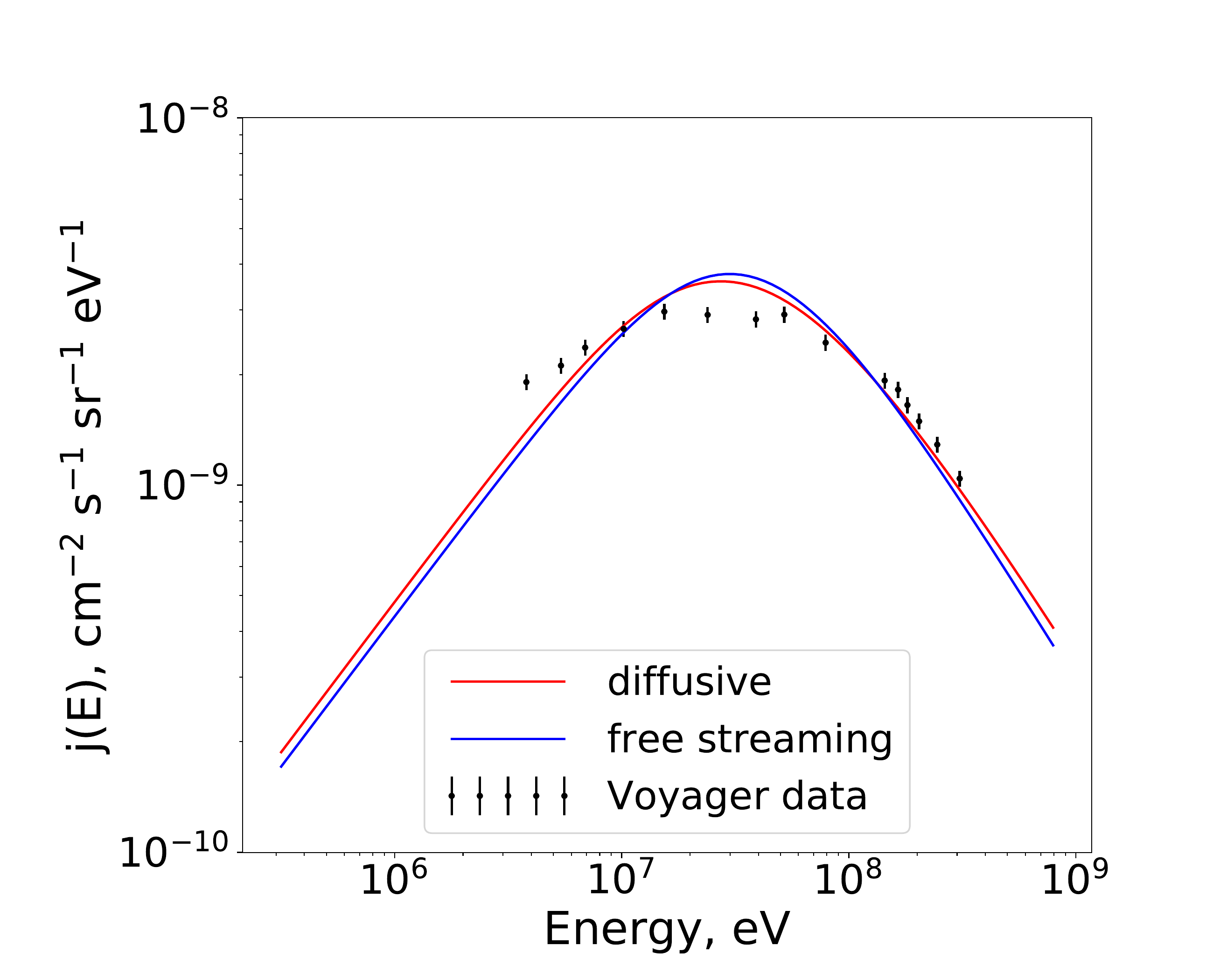}
\caption{The red and blue curves show the best-fit attenuated spectra from Equations \eqref{Jeq} and \eqref{CDSAj} respectively.  The black points with error bars represent the Voyager data \citep{Cummings16}.  }
\label{voyagerFig}
\end{figure}
As mentioned in the previous section, one source of low-energy CRs are shocks in the ISM which are expected to produce a power-law spectrum of accelerated particles.  In particular, in the non-relativistic regime, a strong shock will produce a spectrum of particles with $j(E) \propto E^{-1}$ \citep{Drury83}.  On the other hand, there is evidence \citep{Alves18}, that the local bubble is surrounded by a thin shell of dense material, with the magnetic field nearly in the plane of the shell.  In this picture, CRs penetrating into the local bubble must pass through a significant column density in the shell.
\par
Let us assume the source produces a spectrum of particles outside the shell given by Equation \eqref{initJ} with $a = 1.0$, and $j_0$ a free parameter.  Given a propagation model, then we can find the column density and value of $j_0$ which best fit the Voyager data.  Figure \ref{voyagerFig} shows the best-fit spectra where the column density and the strength of the source spectra are free parameters.  The points are the data from the Voyager probe \citep{Cummings16}.  The blue curve shows the best fit curve assuming free-streaming propagation [Equation \eqref{CDSAj}, integrated over $\mu$], and the green curve shows the best fit assuming diffusive propagation [Equation \eqref{Jeq}].  
\par
 One can see that the best-fit spectra have very similar shapes, although the diffusive propagation model fits the data marginally better.  There is, however, an important difference: in the free-streaming model, the best fit is obtained with a column density of $1.4 \times 10^{23}$ cm$^{-2}$, or $4.7 \times 10^4$ pc/cm$^3$.  Unless the shell is very dense ($>100$ cm$^{-3}$), or the magnetic field extremely close to parallel to the shell (i.e. field lines wrap around the shell multiple times before entering the bubble), it seems difficult to understand from where such a large column could arise.  In the diffusive model, on the contrary, the results depend both on  $N$ and $\Delta_0$.  Assuming $\Delta_0$ to be the same as that used in Section \ref{molCloudIon}, then we find a best fit value of $N$ of $5 \times 10^{21}$ cm$^{-2} = 1.8 \times 10^{3}$ pc/cm$^3$.  This shows that using this model, the required attenuation can occur with a reasonable physical column density.  That said, it is clear that our best fits do deviate significantly from the data, so these simplified models must of course not be the whole story.

\section{Conclusions and Outlook}
We proposed a model for the change in the low-energy CR spectrum (and corresponding ionization rate) as CRs propagate diffusively through a medium (where a certain degree of pre-existing turbulence is present) losing their energy to ionization.  This predicts a substantially steeper slope of the ionization rate $\zeta_{\rm H_2}$ as a function of column density compared with the free-streaming model.  Under conditions appropriate for local molecular clouds, this mechanism would likely only operate up to column densities of $\sim 10^{21}$ cm$^{-2}$.  However, we showed that the assumption of diffusive propagation makes a significant difference to the behavior of $\zeta_{\rm H_2}$, and there are reasonable sets of physical parameters under which it could operate.  We have provided analytic solutions for $\zeta_{\rm H_2}(N)$, Equations \eqref{analDiffusiveZeta} and \eqref{analStreamingZeta}, that can be applied to a variety of environments.
\par
We also considered the question of how the spectrum of CRs seen by Voyager can be produced.  We note that, to produce such a spectrum from a power-law source spectrum (predicted from the theory of diffusive shock acceleration), would require a large attenuating column of $\sim 10^{23}$ cm$^{-2}$.  It is difficult to understand from where this column could arise.  However, if one uses a diffusive propagation model, a marginally better fit to the Voyager spectrum can be obtained while keeping the required column density well under $10^{22}$ cm$^{-2}$.
\par
The principal aim of the present paper is to highlight the stark differences in the behavior of $\zeta_{\rm H_2}(N)$ depending on the mode of CR transport.  More detailed observations and analyses must be performed to distinguish between the two modes.  In particular, it would be desirable to perform a dedicated analysis of $\zeta_{\rm H_2}(N)$ measured in molecular clouds, to determine the most probable slope more reliably.  Furthermore, the analysis of \citet{Neufeld17} should be done \underline{assuming} that $\zeta_{\rm H_2}$ varies within the cloud, rather than assuming a constant value within each cloud.  Also, it would be good to have a more detailed model of CR transport in the shell surrounding the local bubble, based on the current model of the $B$ field, and taking into account transverse diffusion.
\par
\vspace{1 cm}
We would like to thank Marco Padovani and Daniele Galli for useful discussions and suggestions.

\bibliographystyle{apj}
\bibliography{diffusionNote}

\end{document}